\documentclass[granma]{svjour}
\usepackage{graphicx}
\usepackage{epsfig,amssymb,citesort}

\begin{document}

\title{Small and Large Scale Granular Statics}
\author{Chay Goldenberg$^1$ and Isaac Goldhirsch$^2$} 

\institute{ $^1$School of Physics and
  Astronomy\\ Tel-Aviv University\\ Ramat-Aviv, Tel-Aviv 69978, Israel\\
  \email{chayg@post.tau.ac.il} \\ \\
  $^2$Department of Fluid Mechanics and Heat Transfer\\ Faculty of
  Engineering\\ Tel-Aviv University\\ Ramat-Aviv, Tel-Aviv
  69978, Israel\\ \email{isaac@eng.tau.ac.il}\\
  \\ We appreciate helpful interactions with R.~P. Behringer, J. Geng, E.
  Cl\'{e}ment, D.  Serero, H. Jaeger, T.~A. Witten, N. Mueggenburg and J.-N.
  Roux. Support from the U.S.-Israel Binational Science Foundation (BSF),
  INTAS and the Israel Science Foundation (ISF) is gratefully acknowledged.}

\date{Received: \today}
\maketitle

\begin{abstract}
  Recent experimental results on the static or quasistatic response of granular
  materials have been interpreted to suggest the inapplicability of the
  traditional engineering approaches, which are based on elasto-plastic models
  (which are elliptic in nature). Propagating (hyperbolic) or diffusive
  (parabolic) models have been proposed to replace the `old' models.  Since
  several recent experiments were performed on small systems, one should not
  really be surprised that (continuum) elasticity, a macroscopic theory, is not
  directly applicable, and should be replaced by a grain-scale
  (``microscopic'') description.  Such a description concerns the interparticle
  forces, while a macroscopic description is given in terms of the stress
  field.  These descriptions are related, but not equivalent, and the
  distinction is important in interpreting the experimental results. There are
  indications that at least some large scale properties of granular assemblies
  can be described by elasticity, although not necessarily its isotropic
  version. The purely repulsive interparticle forces (in non-cohesive
  materials) may lead to modifications of the contact network upon the
  application of external forces, which may strongly affect the anisotropy of
  the system.  This effect is expected to be small (in non-isostatic systems)
  for small applied forces and for pre-stressed systems (in particular for
  disordered systems).  Otherwise, it may be accounted for using a nonlinear,
  incrementally elastic model, with stress-history dependent elastic moduli.
  Although many features of the experiments may be reproduced using models of
  frictionless particles, results demonstrating the importance of accounting
  for friction are presented.
\end{abstract}
                                
\section{Introduction}
\label{sec:intro}
The modeling of granular materials has been a subject of ongoing research in
the engineering community (see e.g.,~\cite{Levy01}). In recent years, this
subject has found renewed interest among
physicists~\cite{Jaeger96b,Jaeger96a,DeGennes99,Kadanoff99} (having been studied
in the distant past by great physicists such as Coulomb, Faraday, Reynolds and
others).

The behavior of ``granular gases'', which are obtained by e.g., sufficiently
strong shaking or shearing (so that the material behavior is dominated by
interparticle collisions), has been quite successfully modeled using approaches
based on extensions of the kinetic theory of gases \cite{Goldhirsch03}.
However, the behavior of dense granular matter, which is dominated by prolonged
interparticle contact, has proven more difficult for modeling. For the
description of the quasi-static behavior, elasto-plastic models are commonly
used by engineers~\cite{Nedderman92,Savage98b}.

This paper is concerned with the static behavior of granular systems. In
elasto-plastic models, one often uses (linear) elasticity below yield (although
parts of a static system are sometime assumed to be at incipient
yield~\cite{Savage98b}).  However, in recent years a very different class of
models has been proposed for describing the statics of granular materials,
based on the notion of ``force propagation'', suggested by the observation of
force chains in experiments on granular materials~\cite{Drescher72}, as well as
simulations~\cite{Cundall79,Radjai99}.  These models (see
e.g.,~\cite{Wittmer96,Bouchaud95,Wittmer97,Tkachenko99}) typically yield
hyperbolic partial differential equations for the stress field, in contrast
with the elliptic, non-propagating nature of the classical equations of static
elasticity. It has been claimed that the hyperbolic description tends to an
elastic-like one at large scales~\cite{Bouchaud01,Socolar02,Otto03} (however,
the physical interpretation of the macroscopic fields in this case is not
clear).

Recently, the response of granular slabs resting on a horizontal floor to a
`point force' applied at the center of the top of the system has been studied
experimentally~\cite{Silva00,Reydellet01,Serero01,Geng01,Geng03,%
  Mueggenburg02,Moukarzel03}. In~\cite{Silva00,Geng01,Geng03}, the intergrain
force distribution has been measured in two-dimensional (2D) systems as a
function of vertical and horizontal distance from the point of application of
the force.  In~\cite{Moukarzel03}, the particle displacements for similar 2D
systems have been measured. In~\cite{Reydellet01,Serero01,Mueggenburg02}, the
vertical force acting on the floor has been measured in three-dimensional (3D)
systems.  Prominent force chains have been observed in ordered 2D systems;
these force chains fade out with increasing disorder. For pentagonal particles
in 2D arrangements the measured force distribution is single peaked and the
width of the peak is linearly related to the vertical distance, in conformity
with elasticity. The results for cuboidal particles obtained in~\cite{Silva00}
appear to suggest a parabolic behavior, consistent with a diffusive model,
although the systems studied were quite small. In~\cite{Moukarzel03}, the width
of the measured distribution of displacements, as a function of the vertical
distance from the particle which is directly displaced, follows a square root
dependence (as expected from a diffusive model) for small distances of a few
particle diameters, crossing over to a linear dependence at larger distances
(consistent with an elastic description).  Ordered 3D packings exhibit multiple
force peaks for shallow systems~\cite{Mueggenburg02} and less structure for
deeper ones.  Somewhat larger (in terms of number of particles), disordered 3D
systems~\cite{Reydellet01,Serero01} exhibit a single peak in the force
distribution measured at the floor, whose width is proportional to the depth of
the system.

The experimental evidence appears to be contradictory: different experiments
seem to support fundamentally different descriptions of the response of
granular materials (in the case of 2D systems, it has been suggested that there
may be a crossover from a hyperbolic to an elliptic behavior with increasing
disorder~\cite{Geng01}). The thesis presented in this paper is that these
seemingly contradictory experimental results (and theoretical explanations) are
not necessarily at odds with each other.  This thesis is based on the
observation that most of the studies (perhaps all) rejecting the elliptic
description have been devoted to small systems, of the size of a few dozen of
particle diameters at most, whereas many engineering studies consider rather
large granular systems. Since elasticity and other {\em macroscopic}
descriptions are not valid on small scales, at which local anisotropies and
randomness play a major role, one should not be surprised that such
descriptions fail on small scales.  Indeed,
simulations~\cite{Goldenberg02,Goldhirsch02} reveal the existence of a
crossover from microscopic to macroscopic behavior of granular assemblies (as
well as other systems~\cite{Tanguy02}) as a function of system size or
resolution. We argue that such a crossover is observed in some of the
experiments mentioned above.  Strictly isostatic systems~\cite{Moukarzel01}
have been shown to be described by hyperbolic stress
equations~\cite{Tkachenko99,Head01}, and numerical simulations suggest that
systems of frictionless spherical particles approach isostaticity in the limit
of infinite rigidity~\cite{Silbert01}). However, we argue that since real
granular systems have finite rigidity and usually experience frictional
interactions, they cannot be generically isostatic (the same presumably holds
even for frictionless non-spherical grains). The isostatic limit is a singular
case, whose physical consequences for {\em real} systems are at best unclear.
Therefore the controversy surrounding the correct description of granular
statics is mostly a question concerning the behavior of small granular systems.
The latter require a grain-scale (``microscopic'') description, rather than a
macroscopic one.

A second point stressed below is the distinction between {\em force} and {\em
  stress}.  Whereas interparticle forces can exhibit force chains which look
like they contradict elasticity, the latter does not describe the nature of the
forces but rather that of the stress field. The stress field involves an
averaging over the forces (whose result is resolution dependent) and leads to
less pronounced structure than the underlying force field. The small scale
structure of the interparticle forces cannot be taken to consist an argument
against an elliptic description or in favor of it, since it relates to small
scales and it does not deal with the objects with which elasticity or
plasticity are concerned. The large scale response of granular packing is shown
to be consistent with a (possibly anisotropic) elastic description. The fact
that in non-cohesive granular materials there is no significant attraction
among the particles may lead to modifications of the contact network, which may
strongly affect the anisotropy of the system. This effect is expected to be
small for small applied forces (for non-isostatic systems) and for pre-stressed
systems, in particular for disordered systems. Otherwise, it may be accounted
for using a nonlinear, incrementally elastic model, with stress-history
dependent elastic moduli.

The third point made in this paper is that while models employing frictionless
particles can reproduce some properties of granular packings, friction can be
of utmost importance for the description of granular matter (a rather intuitive
fact).  Results demonstrating the importance of accounting for frictional
interactions are presented in Sec.~\ref{sec:friction}.

\section{The microscopic picture: forces}
\label{sec:micro_forces}

In attempting to describe granular materials in terms of continuum mechanics,
by analogy to ``regular'', atomic materials, one usually considers the
``microscopic'' scale to be that of the individual particles (whose internal
dynamics should be well described by continuum mechanics).

One of the simplest granular systems is a collection of frictionless spherical
particles. A typical microscopic (particle scale) description of such a system
is given by the particle's radii, $\{R_i\}$, their masses, $\{m_i\}$, center of
mass positions, $\{\vec{r}_i(t)\}$, and velocities, $\{\vec{v}_i(t)\}$, at time
$t$.  It is typically assumed (e.g., in the context of simulations of granular
materials~\cite{Cundall79,Herrmann98,Wolf96}) that the particles are quite
rigid, so that the interaction between two particles (in the frictionless case)
depends only on their respective distance, or, more conveniently, on their
imaginary overlap $ \xi_{ij}(t)\equiv R_i+R_j-|\vec{r}_{ij}(t)|$, where
\mbox{$\vec{r}_{ij}(t) \equiv \vec{r}_{i}(t)-\vec{r}_{j}(t)$}. The contact
interactions are usually modeled by treating the particles as macroscopic
objects, described by the equations of continuum mechanics (see
e.g.,~\cite{Gladwell80,Johnson85}). For two frictionless elastic spheres, a
classical result by Hertz (see e.g.,~\cite{Landau86}) is that the force is
proportional to $\xi^{3/2}$, while for cylinders, it is linear in the overlap.
For noncohesive particles, only repulsive forces are possible.  Even for
frictionless particles, internal dissipation as in e.g., viscoelastic
particles, gives rise to a dependence of the force on the relative velocity
$\dot{\xi}$ as well (for some examples of force schemes commonly used in
simulations, see e.g.,~\cite{Schafer96,Sadd93,Walton95}).

The interparticle forces for a given configuration of such particles subject to
given boundary conditions (e.g., specified displacements of the particles on
the boundary, or forces applied to them) and body forces such as gravity can be
determined, for a static system, using the equations of equilibrium (Newton's
laws) and the force-displacement relation. We reiterate that when full force
laws for particle interactions are known or modeled, the statics and dynamics
of the system are fully determined (they may be history-dependent for
history-dependent force laws, as commonly used for frictional interactions).

In the case of frictionless isostatic systems (in which the mean coordination
number is {\em exactly} $z=2d$, where $d$ is the dimension of the system) the
forces can be determined from the equations of equilibrium alone (and are
therefore independent of the force-displacement law; however, the particle {\em
  displacements} certainly depend on this law). It has been
suggested~\cite{Moukarzel98c,Roux00} that frictionless granular systems become
isostatic in the limit of infinite rigidity (giving rise to a macroscopic
behavior which is very different from elasticity), and this appears to be borne
out by numerical simulations~\cite{Makse00,Silbert01}. However, the relevance
of this limit to real materials is questionable, since real materials cannot be
infinitely rigid. Any additional contacts created if the rigidity is allowed to
be finite will render the system hyperstatic (so that there is a ``phase
transition'' to an isostatic behavior {\em only} at infinite
rigidity~\cite{Moukarzel98c}).  The rigidity should of course be compared to
the confining forces or body forces (in a system under gravity and confined by
walls, the confining force is related to gravity). If the confining forces are
very small, the system would indeed be expected to be close to marginal
stability.  As mentioned above, the static indeterminacy associated with
hyperstatic systems simply means that the equations of equilibrium are
insufficient for determining the forces, so that additional equations (e.g.,
force-displacement laws) are required.  Static indeterminacy {\em does not}
mean that there's no unique solution for the forces in a {\em real} system. A
similar situation occurs on the macroscopic, continuum level (see
e.g.,~\cite{Savage98b}). The rigid limit can be approached in many different
ways (e.g., the stiffness of each interparticle contact may be different), and,
even if assuming that the same (isostatic) contact network is obtained for
different distribution of the interparticle stiffness, yielding the same
interparticle forces, the particle displacements will certainly be different,
hence the rigid limit in not unique, at least in this sense.

In several experiments, photoelastic particles were used in order to measure
the stress in granular
systems~\cite{Drescher72,Howell99,Silva00,Geng01,Geng03}.  These measurements
probe the {\em intraparticle} stress, i.e., the stress {\em inside} each
particle.  Following the above, these should be interpreted as measurements of
microscopic fields (the macroscopic description of granular systems regards the
particles as microscopic, and does not resolve any details below the particle
scale). The microscopic fields corresponding to these measurements are the
interparticle forces, which can be deduced from these internal stress
measurements (as described in~\cite{Geng03}). As mentioned, these forces should
be distinguished from the ``macroscopic'' stress field in the system.

The distribution of force magnitudes in a static granular packings is a
microscopic quantity which has been extensively studied in
experiments~\cite{Mueth98,Blair01} and simulations~\cite{Radjai96}. An
exponential behavior of the distribution at large forces appears to be quite
universal in experiments on granular systems, independent of the degree of
disorder~\cite{Blair01}, the friction coefficient~\cite{Blair01}, or the
rigidity of the particles~\cite{Erikson02}, and has also been observed in
simulations of granular systems with different models for the interparticle
forces (e.g.,~\cite{Nguyen00,Silbert02a}). The universality of the force
distribution appears to extend to other systems such as foams, glasses,
colloids etc.  (see~\cite{O'hern00} and references therein). The exponential
tail of the distribution is reproduced in simple models such as the (parabolic)
q-model~\cite{Liu95,Coppersmith96}. The distribution for smaller forces
appears to be less universal, and it has been suggested that the appearance of
a peak in the force distribution near the mean force may signal the onset of
jamming or a glass transition~\cite{O'hern02}.

\begin{figure}
  \centerline{\includegraphics[width=\hsize,clip]{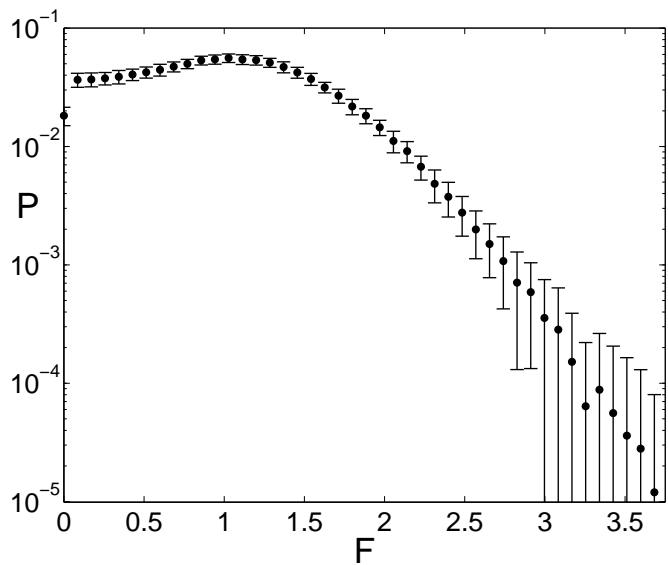}}
   \caption{The distribution of force magnitudes in bond-diluted distorted
     triangular networks of linear springs (see text).\label{fig:forcedist}}
\end{figure}

Interestingly, a qualitatively similar force distribution is obtained in purely
harmonic networks: Fig.~\ref{fig:forcedist} shows the force distribution
obtained for an ensemble of random networks of linear springs constructed as
follows. Points are placed on a 2D triangular lattice with spacing $d$ (with
square-shaped boundaries), and then their $x$ and $y$ coordinates are randomly
displaced by $\pm 0.04d$. Points whose distance is less than $1.02d$ are
connected by linear springs (whose equilibrium length is equal to this
distance) with equal spring constants (this results in an average dilution of
about $12\%$ of the springs compared to the perfect lattice).  A uniform
isotropic compression of $1\%$ is applied to the boundary particles, and the
interparticle forces are calculated. The force distribution presented in
Fig.~\ref{fig:forcedist} is obtained from an average over the force histograms
of $100$ systems of $1085$ particles. The force was normalized by the mean
force in the ensemble (a similar distribution is obtained for a normalization
by the mean force for each system; the variation in mean force among different
systems is relatively small, which may indicate that the system is far from
`jamming'~\cite{O'hern02}). The tail of the logarithm of the distribution is
fit quite well with a line of slope $-3.8$, similar to the slope obtained in
experiments on highly compressed disordered packings of soft rubber
spheres~\cite{Erikson02} (similar distributions were obtained for a scalar
harmonic network of unequal springs in~\cite{Nguyen00}).  For the case of
networks with no force dilution (the same connectivity as in the perfect
lattice), the force distribution is Gaussian with a half-width of a few percent
of the mean, i.e., a much narrower distribution). These results indicate that a
random connectivity should be consequential for the force distribution, which
may be the reason that even for highly compressed disordered spheres (whose
contact network is still disordered), the distribution is qualitatively similar
to that observed in less compressed systems~\cite{Erikson02}. A similar effect
has been observed in simulations of granular systems under different applied
pressures~\cite{Nguyen00}.

\begin{figure}
  \centerline{\includegraphics[width=\hsize,clip]{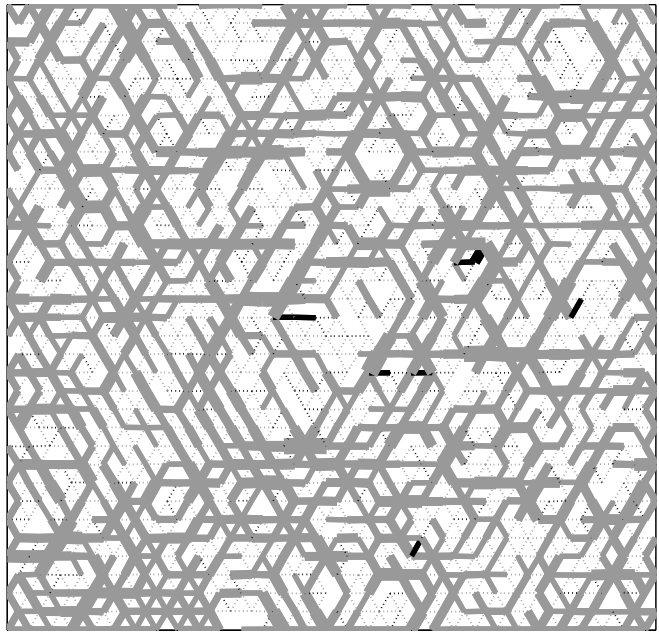}}
   \caption{The forces in a bond-diluted distorted
     triangular network of linear springs. Forces with magnitude larger than
     the mean are indicated by solid lines whose width is proportional to the
     force magnitude; smaller forces are indicated by thin dotted lines.
     Compressive forces are indicated by gray lines; tensile forces by
     black lines.\label{fig:forcechains_diluted}}
\end{figure}

The forces in one of the realizations of the ensemble are shown in
Fig.~\ref{fig:forcechains_diluted}.  Force chains are clearly observed (note
that there are very few tensional forces, so that they do not significantly
affect the force distribution in this case). Similar force chains have been
observed in a polydisperse Lenard-Jones system~\cite{Tanguy02} (incidentally,
the concept of a force chain is not well-defined: in the case of a homogeneous
strain applied to a uniform lattice, the forces are equal, so that it is
reasonable to define the force chains to contain forces whose magnitude is
larger than a uniform cutoff, e.g., the mean force, as used in
Fig.~\ref{fig:forcechains_diluted}; However, for a non-uniformly strained
system, e.g., systems subject to gravity, in which the mean force increases
with depth, such a global cutoff makes little sense).  The results shown in
Fig.~\ref{fig:forcechains_diluted} indicate that force chains are not specific
to granular systems.  Force chains are {\em microscopic} features of
microscopically disordered systems (or even inhomogeneously strained ordered
systems, as described below), and their presence does not necessarily indicate
any macroscopic inhomogeneity, or inconsistency with a macroscopic elliptic, or
elastic, description. It is quite certain that if one could observe the
individual interparticle forces in atomic systems (which may not be quite well
defined, since a quantum description is appropriate for such systems), one
would also observe force chains.

\begin{figure*}
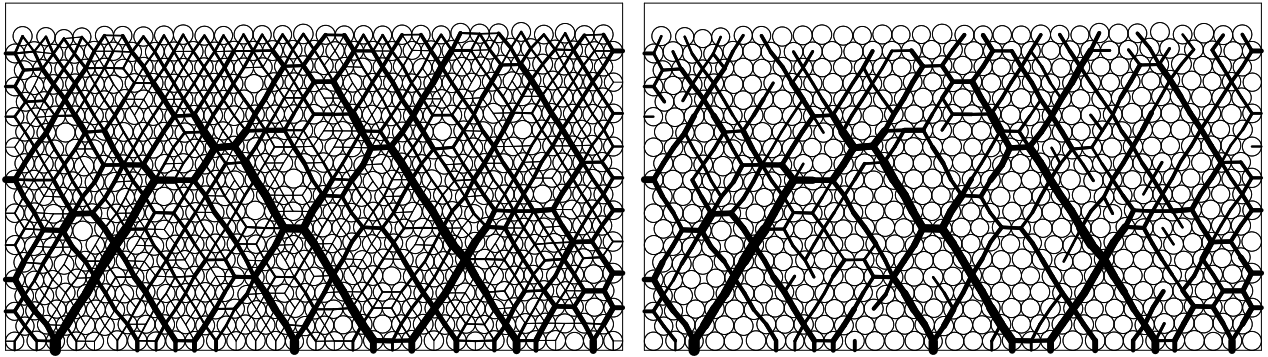

  \begin{tabular}{cc}
&\includegraphics[width=3.3in,clip]{chops04_gm070_v2_fig3.eps}
\includegraphics[width=3.3in,clip]{chops04_gm070_v2_fig4.eps}
  \end{tabular}
   \caption{The forces in a system of polydisperse frictionless disks with a
     uniform force applied to the top layer (no gravity). Line widths are
     proportional to the force magnitudes. Left: all forces, right: only the
     forces whose magnitude is larger than the
     mean.\label{fig:forces_all_gtmean}}
\end{figure*}

\begin{figure}
  \centerline{\includegraphics[width=\hsize,clip]{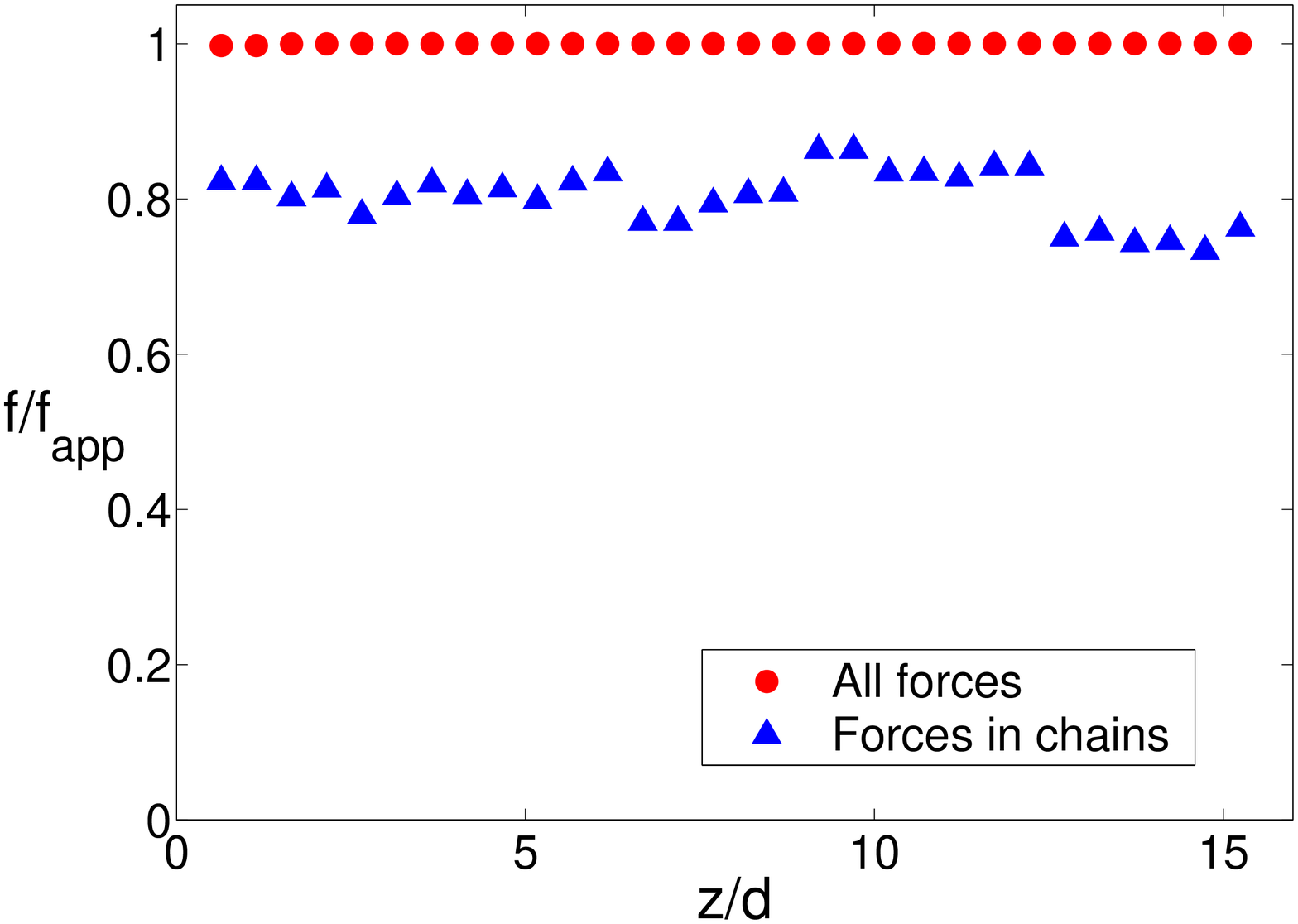}}
   \caption{The fraction of the applied vertical force carried by all the forces and by
     the forces whose magnitude is larger than the mean (i.e., those belonging
     to force chains), as a function of the vertical coordinate, $z$, scaled by
     the mean particle diameter, $d$, for the system shown in
     Fig.~\ref{fig:forces_all_gtmean}.\label{fig:forcechainspercent}}
\end{figure}

It is important to note that a significant portion of the stress (even in a
homogeneously strained system) is carried by forces which do not belong to the
force chains. An example is provided by a system of frictionless polydisperse
disks (with radii uniformly distributed within $10\%$ of the maximum radius)
which is confined by side walls and a floor, with a uniform force applied to
the particles of the ``top'' layer (without gravity). The interparticle forces
are taken to be linear in the overlaps. Fig.~\ref{fig:forces_all_gtmean} shows
the forces in the system. Fig.~\ref{fig:forcechainspercent} shows the fraction
of the applied vertical force carried by the forces whose magnitude is greater
than the mean (i.e., those belonging to force chains, using the definition
mentioned above), compared to that carried by all the forces (which is of
course equal to $1$), for forces in horizontal ``slices'' of the system, as a
function of the vertical coordinate, $z$. As seen in
Fig.~\ref{fig:forcechainspercent}, only about $80\%$ of the applied force is
carried by the force chains. Furthermore, the force carried by the chains
fluctuates with depth, so that the forces in the chains do not obey the
conditions of force equilibrium. It is therefore questionable whether a model
which describes the stress exclusively in terms of the force chains is
justifiable.

The (near-)universality of the force distribution, in particular the fact that
it is observed in simulations of random systems with harmonic interactions,
does not make possible the differentiation between different models on the
basis of the force distribution (in particular, the observation of such a
distribution does not preclude an elliptic description). The same statement
applies to the observation of force chains. A more sensitive and direct test
should be rendered by the response of a granular system to inhomogeneous
external forcing, such as that provided by localized forces. The latter seem to
be consistent with elasticity, as described below.
\section{Macroscopic fields and continuum equations  in terms of microscopic
  quantities}
\label{sec:cont_eqs}

Continuum descriptions of materials are often based on phenomenological
arguments (usually motivated by experimental findings), rather than on
derivations from the underlying microscopic dynamics. A unique feature of
granular materials is that due to the typically large sizes of the
constituents, it is relatively easy to access the ``microscopic'' scales
experimentally. On the other hand, in most practical applications, the number
of particles is such that a detailed particle-level description becomes
intractable, and a continuum description is required. The fact that experiments
on granular systems can yield both microscopic information and macroscopic
information (possibly even in the same experiment) is useful to the elucidation
of the connection between these two descriptions.


In order to obtain a macroscopic description of a system in terms of the
microscopic fields, we employ a spatial coarse-graining
approach~\cite{Glasser01,Goldhirsch02}. We stress that the only averaging
considered here is spatial (the approach can be extended to include temporal
coarse-graining as well~\cite{Glasser01}, but here we consider static
configurations). Since static granular packings are typically found in
metastable states, far from equilibrium, and thermal energy scales are
negligible, such systems do not explore any phase space so that it is hard to
justify the kind of ensemble average commonly used in statistical mechanics. An
average over configurations (i.e., average over different disordered systems
which are presumed to be prepared in the same way) is commonly performed when
analyzing experimental data, due to the large fluctuations obtained in many
experiments. However it is not clear a-priori if self-averaging occurs, i.e.,
that at least for large enough scales the macroscopic behavior of a single
``typical'' realization is the same as that of the average behavior over many
realizations. Self-averaging may be valid for some quantities and not for
others.  Therefore we choose not to assume a-priory that any ensemble averaging
is justified; instead we relate the macroscopic and microscopic fields in a way
that is relevant for single realizations.

Following~\cite{Glasser01}, define the coarse-grained (CG) mass density
$\rho(\vec{r},t)$ and momentum density $\vec{p}(\vec{r},t)$ at position
$\vec{r}$ and time $t$ as
\begin{eqnarray}
  \label{eq:cg_density_momentum}
\rho(\vec{r},t)&\equiv& \sum_i m_i\phi[\vec{r}-\vec{r}_i(t)],  \\
\vec{p}(\vec{r},t)&\equiv& \sum_i m_i \vec{v}_i(t) \phi[\vec{r}-\vec{r}_i(t)],
\end{eqnarray}
where $\phi(\vec{R})$ is a non-negative coarse-graining function (with a single
maximum at $\vec{R}=0$) of width $w$, the coarse-graining scale, and
$\int\phi(\vec{R})d\vec{R}=1$.

Upon taking the time derivative of the macroscopic fields $\rho$ and $\vec{p}$,
performing straightforward algebraic manipulations~\cite{Glasser01} and using
Newton's laws, one obtains the equation of continuity and the momentum
conservation equation, respectively:
\begin{eqnarray}
  \label{eq:cg_conservation_equations}
\dot{\rho} &=& -{\rm div}(\rho\vec{V})\\
\dot{p}_\alpha &=& - \sum_{\beta} \frac{\partial}{\partial
  r_\beta} \left[ \rho V_{\alpha} V_{\beta} -
  \sigma_{\alpha\beta}\right],\nonumber
\end{eqnarray}
where the velocity field is defined by \mbox{$\vec{V}\equiv \vec{p}/\rho$},
Greek indices denote Cartesian coordinates, and the explicit dependence of the
CG fields on $\vec{r}$ and $t$ has been omitted for compactness. Since this
paper focuses on the stress field, we have omitted the energy equation, which
can be derived in a similar way~\cite{Goldhirsch02}.

In addition to obtaining the standard equations of continuum mechanics from
microscopic consideration, this coarse graining procedure provides an
expression for the stress tensor $\sigma_{\alpha\beta}$ in terms of 
the microscopic entities:

\begin{eqnarray}
\label{eq:stress}
\sigma_{\alpha\beta}(t) &=&
-\sum _ {i} \, m_i\, 
v^{\prime} _{i\alpha}   (\vec{r},  t ) \;   v^{\prime} _ {i\beta}  
(\vec{r}, t)   \phi   (\vec{r} - \vec{r}_{i} (t) ) \\
&&-     \frac{1}{2} \sum_{ij;i\ne j} f_{ij\alpha}(t) {r}_{ij\beta}(t) \int_0^1 ds 
\phi[\vec{r} - \vec{r}_{i} (t) + s \vec{r}_{ij}(t)],\nonumber 
\end{eqnarray}
where $\vec{v}'_i(\vec{r},t) \equiv \vec{v}_i(t) - \vec{v}(\vec{r},t)$ is the
fluctuating velocity, $\vec{f}_{ij}(t)$ is the force exerted on particle $i$ by
particle $j$, and $\vec{r}_{ij}(t)\equiv \vec{r}_{i}(t)-\vec{r}_{j}(t)$.

The first term in Eq.~(\ref{eq:stress}) is the kinetic stress (which vanishes
for static configurations), and the second term is known as the contact stress.
Note that the standard Born-Huang expression~\cite{Born88}:
$\sigma_{\alpha\beta} = - \frac{1}{2V} \sum_{ij\in V;i\ne j} f_{ij\alpha}
{r}_{ij\beta}$ is equivalent to the expression for the contact stress in
Eq.~(\ref{eq:stress}) if the coarse-graining function is taken constant inside
a volume $V$ and zero outside it, provided that the interparticle separation is
much smaller than the coarse-graining length scale (typically $\sqrt[3]{V}$).

The above expressions can be used to calculate the macroscopic fields from the
microscopic ones (obtained e.g., from simulations or experiments), and compare
them to the predictions of macroscopic models or direct experimental results.
In order to close the set of continuum equations
[Eqs.~(\ref{eq:cg_conservation_equations})] the stress and energy flux (the
latter is not related to the considerations below) need to be expressed as
functionals of the pertinent {\em macroscopic} fields.  As mentioned, such
constitutive relations are often obtained empirically or conjectured. In some
cases they are derived from the microscopic dynamics. The above {\em exact}
expression for the stress field provides a framework for a systematic
derivation of constitutive relations (as suggested for elastic networks
in~\cite{Goldhirsch02}).

Here, we are concerned with the interpretation of experimental data in terms of
microscopic variables and macroscopic fields.  The fact that the contact stress
includes a sum over all contacts for each particle (i.e., even for very small
CG scales the stress components correspond to specific ``averages'' over the
forces on each particle) already suggests that a ``picture'' of the forces in
the packing does not correspond directly to the macroscopic stress field (they
are certainly related, i.e., one would usually expect large stress components
in regions where large force magnitudes are observed).  In particular, as shown
in Sec.~\ref{sec:micro_forces}, and further discussed below, force chains do
not necessarily indicate macroscopic anisotropy or inhomogeneity.

\section{Numerical results for model frictionless systems}
\label{sec:frictionless}

\begin{figure}
  {\centerline{\includegraphics[width=\hsize,clip]{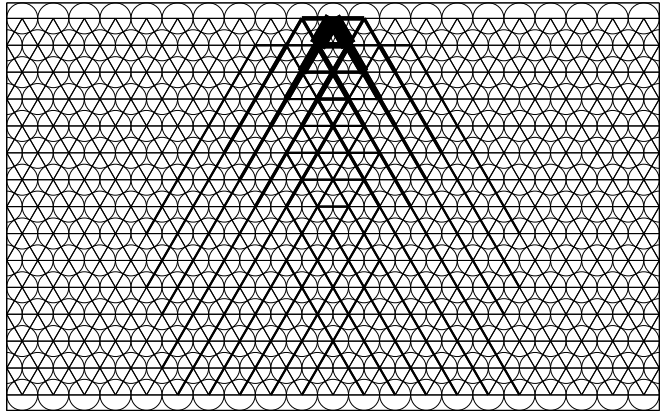}}
   \caption{Force chains in a 2D triangular lattice. A vertical
     force is applied at the center of the top layer. Line widths are
     proportional to the force magnitudes. Only the central part of the system
     is shown; reproduced
     from~\cite{Goldenberg02}.\label{fig:forcechains_harmonic}}}
\end{figure}

Consider a two-dimensional system of uniform disks (arranged on a triangular
lattice) subject to a vertical external force at the center of the top
layer~\cite{Goldenberg02}. Experiments on such systems are described
in~\cite{Geng01,Geng03}. Consider first the case of nearest neighbor harmonic
interactions, i.e., the disks are coupled by equal linear springs (whose rest
length is the diameter of a disk). Clearly, real cohesionless particles do not
experience any significant attractive interactions; however, there are a few
insights to be obtained from the study of this system.
Fig.~\ref{fig:forcechains_harmonic} presents the forces in the system.  Force
chains are evident.

\begin{figure}
  {\centerline{\includegraphics[width=\hsize,clip]{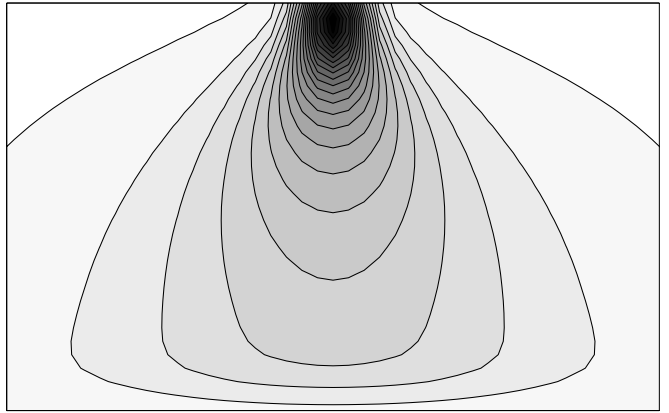}}
    \caption{Contour plot of $h\sigma_{zz}$, corresponding to
      Fig.~\ref{fig:forcechains_harmonic} ($h$ is the slab
      height); reproduced from~\cite{Goldenberg02}.\label{fig:stress_twosided_contour}}}
\end{figure}

A contour plot of the ``vertical stress component'' $\sigma_{zz}$ [computed
using Eq.~(\ref{eq:stress})] for the same system is shown in
Fig.~\ref{fig:stress_twosided_contour} (with $\phi(\mathbf{r})=\frac{1}{\pi
  w^2}e^{-(|\mathbf{r}|/w)^2}$, and \mbox{$w=d$}, the particle diameter, i.e.,
a fine resolution). The force chains are not evident any more. The model
described above corresponds, in the continuum (long-wavelength) limit, to an
isotropic 2D elastic medium~\cite{GoldenbergUP}. The observed force chains,
which break isotropy, can be attributed to the fact that the local environment
of a particle in contact with a finite number of other particles cannot be
isotropic. Under homogeneous macroscopic deformation, all forces would be equal
in a lattice configuration. However, the concentrated applied force yields an
inhomogeneous deformation, which leads to the local anisotropy being reflected
in the distribution of the forces.  The elastic continuum description of the
stress (to linear order in the strain) is isotropic, and cannot be expected to
reflect this microscopic anisotropy. For small system sizes (in which the
strain gradients on a particle scale are relatively large), this anisotropy can
be observed in the stress field (a very clear example is shown
in~\cite{Goldenberg02} for a macroscopically isotropic 3D system, whose
microscopic symmetry is cubic).  These results, as well as those presented in
Sec.~\ref{sec:intro} for disordered elastic systems, show that force chains do
not necessarily indicate anisotropy or inhomogeneity of the material on
sufficiently large scales; more importantly their existence does not require a
non-elastic (microscopic) interaction.

Note that only the forces between the particles and the floor (a single such
force per particle) are used in the calculation of the stress at the bottom of
the packing. Hence on the bottom (but not in the bulk of the system), the
spatial distribution of $\sigma_{zz}$ is equivalent (up to coarse-graining) to
that of the microscopic forces.  For sufficiently large systems, the
distribution of forces on the bottom corresponds closely to the stress
calculated using linear elasticity~\cite{Goldenberg02}, even ``almost without
coarse-graining'', i.e., for a microscopic CG scale.

\begin{figure}
  {\centerline{\includegraphics[width=\hsize,clip]{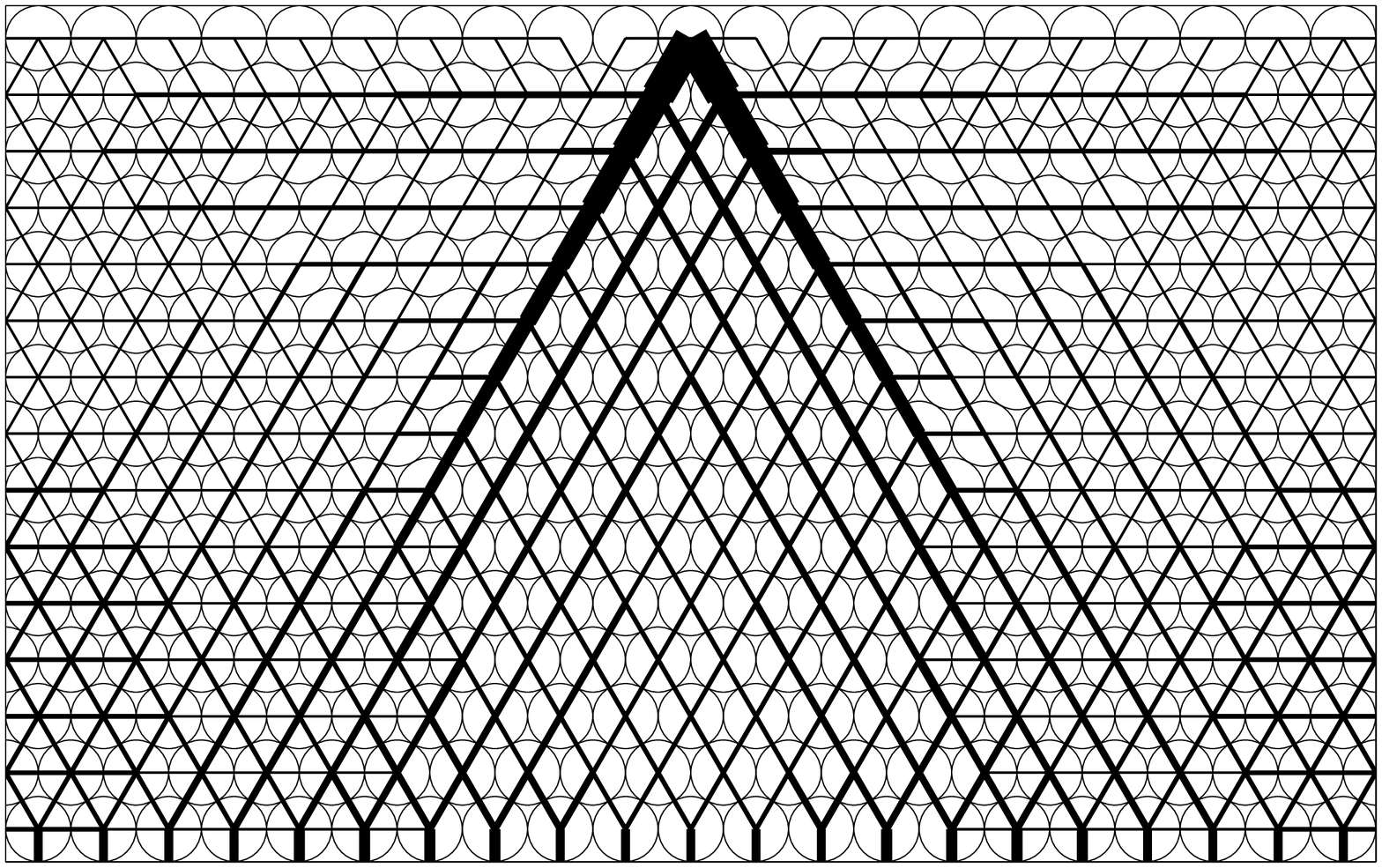}}
    \caption{Force chains in a 2D triangular lattice of  `one-sided'
      springs. A gravitational force has been applied in order to stabilize the
      system (the applied force is 150 times the particle weight); reproduced
      from~\cite{Goldenberg02}.\label{fig:forcechains_onesided}}}
\end{figure}

A more realistic force model consists of `one-sided' springs, i.e., springs
that snap when in tension.  Fig.~\ref{fig:forcechains_onesided} presents the
forces obtained for the same system presented in
Fig.~\ref{fig:forcechains_onesided}, but with `one-sided' springs. Compared to
the system of regular springs, the application of the concentrated force at the
top of the packing leads to rearrangements in the contact network: some
horizontal springs in the region under the point where the force is applied are
disconnected (as also observed in~\cite{Luding97} for a pile geometry) but the
force chains in both systems are qualitatively similar.  The force distribution
vs. the horizontal coordinate at different depths is in good
agreement~\cite{Goldenberg02} with experiment~\cite{Geng01,Geng03}.  For slightly
disordered systems~\cite{Goldenberg02}, the force chains are qualitatively
similar, though somewhat shorter.

\begin{figure}
  {\centerline{\includegraphics[width=\hsize,clip,clip]{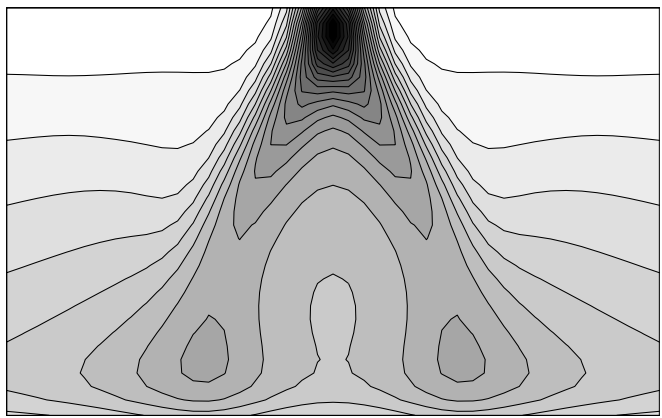}}
    \caption{Same as Fig.~\ref{fig:stress_twosided_contour}, for the
      case of `one-sided' springs (for which the forces are shown in
      Fig.~\ref{fig:forcechains_onesided}); reproduced from~\cite{Goldenberg02}.
      \label{fig:stress_onesided_contour}}}
\end{figure}

The corresponding vertical stress field $\sigma_{zz}$ is shown in
Fig.~\ref{fig:stress_onesided_contour}. The stress field in this case is
clearly quite different from that obtained using `two-sided', harmonic springs:
the response for `one-sided' springs is double-peaked. This is obviously
related to the disconnected springs below the point of application of the
external force. In~\cite{Goldenberg02}, it has been shown that a model with
harmonic springs in which the spring constant for the horizontal springs,
$K_1$, is different from that of the oblique springs, $K_2$, corresponds (in
the continuum limit) to an anisotropic elastic system. For sufficiently large
$K_2/K_1$, the response of such an elastic system has two
peaks~\cite{Goldenberg02} (see also~\cite{Otto03} for a more detailed analysis
of the case of an infinite half-plane; the results presented
in~\cite{Goldenberg02} are for a finite slab on a rigid floor). The absence of
horizontal springs corresponds to the limit $K_2/K_1\rightarrow \infty$, the
extreme anisotropic limit, which corresponds to an isostatic system. Note that
the {\em stress} field, but not the displacement, depends only on $K_2/K_1$.
In the case considered here, $K_1=0$ and $K_2$ is finite; for $K_2\rightarrow
\infty$, the rigid limit, the displacement is zero. The double peaked stress
distributions are similar to those obtained from hyperbolic models.  It follows
that hyperbolic-like behavior can be obtained using an \textit{anisotropic}
(yet, still elliptic) elastic model (which becomes formally `hyperbolic' in the
limit of very large anisotropy; see also~\cite{Cates99,Otto03}).

The `stress-induced anisotropy'~\cite{Goldenberg02} observed in the case of
`one-sided' springs can be thought of as a nonlinear extension of the linear
elastic continuum behavior obtained in a network of harmonic springs. While the
(possibly position-dependent) elastic moduli in linear elasticity are
time-independent material properties, a possible extension would be to
introduce a stress history dependence of the elastic moduli (i.e., the
anisotropy induced by the breaking of contacts in certain regions may be
considered a result of a tensile stress in those regions).  A similar type of
stress-induced anisotropy has been suggested in the context of plastic models
for soil mechanics~\cite{Oda93}. If the particle positions do not change
significantly, so that only the contact network is modified in response to the
applied stress, the behavior can possibly modeled as `incrementally elastic'.
Under certain condition (corresponding to plastic yield), the system would no
longer be able to support the applied stress without a major rearrangement of
the particles. Incipient plastic yield may possibly be related to a local
extreme anisotropy typical of a marginally stable isostatic configuration.

The force chains obtained both for the harmonic case and the `one-sided' case
are quite similar~\cite{Goldenberg02} to those observed
experimentally~\cite{Geng01,Geng03}, as an average over different realizations.
This averaging is required due to experimental variations: although the
particles are arranged on lattice, there is still some disorder present due to
some variability in particle diameter, and possibly also in the contact
properties~\cite{GengPC}. Indeed, while perfect atomic lattices may be obtained
at low enough temperatures, since all atoms of the same isotope are exactly
identical, macroscopic particles are never truly identical, so that perfect
periodicity can never be obtained. It also appears that the force chains
obtained using the two models are quite similar (similar chains are also
obtained in slightly disordered systems~\cite{Goldenberg02}).  The stress field
appears to be more sensitive to the anisotropy induced by the applied force.

In the experiments reported in~\cite{Reydellet01,Serero01,Mueggenburg02}, the
forces on the floor were measured.  In~\cite{Reydellet01,Serero01}, the width
of the pressure probe (which would correspond to the CG scale of the measured
stress) was $10-30$ particle diameters. The bottom stress profiles measured are
quite consistent with continuum elasticity (note that the depths of the systems
studied were $20-300$ particle diameters).  Experimental deviations from the
predictions of {\em isotropic} elasticity~\cite{Serero01} can be reproduced by
anisotropic elasticity~\cite{Goldenberg02}. Narrower or wider peaks than those
obtained for isotropic systems can be obtained for small anisotropy, while for
very large anisotropy, two peaks are expected (see also~\cite{Otto03}). An
additional possible cause for deviations from the isotropic elastic
calculations presented in~\cite{Serero01} is finite rigidity of the
floor~\cite{GoldenbergUP}. The more shallow systems used in the experiments may
even be small enough for the finite size effects~\cite{Goldenberg02} to be
significant. Any anisotropy in these experiments is obviously much weaker than
the strong anisotropy observed in the model ordered system of `one-sided'
springs. Several effects may explain this: first, the systems used in the
experiment are highly disordered, so that inhomogeneous, random anisotropy may
be expected on intermediate (already macroscopic) scales, presumably averaging
out to an isotropic, or nearly isotropic, behavior at sufficiently large
scales. In this case, the large-scale effect of contacts breaking due to
applied forces would be significantly less pronounced than in the ordered
system described above. A second possibility is the effect of frictional
forces, which may either prevent contacts from breaking, or reduce the
anisotropy of the response. Third, the model systems discussed above were
unstressed before the application of the force, while the experimental ones are
pre-stressed by gravity, which may compress some of the contacts such that the
tension due to the applied force is insufficient to break them.

In~\cite{Mueggenburg02}, individual forces on the floor were measured, and the
results were averaged over realizations (which, as mentioned, is not
necessarily equivalent to spatial coarse-graining). The regular packings used
in~\cite{Mueggenburg02} (FCC and HCP) are macroscopically anisotropic. The fact
that some of the horizontal contacts (contacts among particles in the same
layer) may be absent increases the anisotropy further (possibly in an
inhomogeneous way; as mentioned above, a granular packing cannot be perfectly
periodic).  The extreme limit in which there are no such horizontal contacts
corresponds to an isostatic system.  Such anisotropy (possibly further enhanced
by the applied force) may explain the discrete peaks observed for relatively
shallow systems composed of $9$ layers of particles (and the fact that they
appear to be consistent with a picture of ``force propagation'' appropriate for
isostatic systems). However, for deeper system (about $20$ particle diameters),
there appears to be a crossover to a smoother behavior, which should correspond
to the crossover to the continuum limit (note that the depth of the systems
used in~\cite{Mueggenburg02} was smaller than the depth required in our
calculations on 3D systems~\cite{Goldenberg02} for reaching the continuum
limit, so deeper systems may still show dependence on the depth).

\section{Effects of Friction}
\label{sec:friction}
As shown, some features of granular response may be reproduced using models
employing frictionless and even harmonically interacting particles. However, it
is clear that friction is consequential for granular materials.

For frictional spheres, the microscopic description, as described in
Sec.~\ref{sec:micro_forces}, must be extended to include (at least) the
orientations of the particles, and interparticle torques in addition to the
forces. The description of static and kinetic friction requires the use of more
complicated force models, which depend on the particle orientations and their
relative tangential velocities, and possibly on the history of contact
deformation (see e.g.,~\cite{Schafer96,Sadd93,Walton95,Vu-Quoc99}).
 
In experiments performed on regular 2D packings of photoelastic
disks~\cite{Geng03}, the directions and ``strengths'' of the force chains
observed upon application of a localized force to the top of the packing appear
to depend quite strongly on the angle of the applied force with respect to the
horizontal (in the following, all angles are given with respect to the
horizontal). A particularly intriguing effect is that for some angles, force
chains appear not only in the lattice directions ($0,\pm 60^{\circ}, \pm
120^{\circ}, 180^{\circ}$ for a triangular lattice), but also, apparently, in
new directions which can be identified as $\pm 30^{\circ}$ (in fact, in
individual realizations, rather than their average as reported
in~\cite{Geng03}, it appears that force chains appear also at $\pm 90^{\circ}$,
i.e., the vertical direction~\cite{GengPC}). These directions correspond to
next-nearest neighbor directions in the triangular lattice.  Since interactions
among the particles only exist for particles in contact, there is no direct
next-nearest neighbor interaction. The fact that the forces themselves, and not
just the contact points, appear to be aligned with these $\pm 30^{\circ}$
directions, suggests that frictional forces among the particles (tangential to
the contact normals, which result in interparticle torques) are necessary for
obtaining forces (and chains) at angles different from the lattice directions.
For an applied force at $\pm 90^{\circ}$, it appears that the frictional forces
are small enough such that the results obtained in this
case~\cite{Geng01,Geng03} are described quite well by a model of frictionless
particles with linear force-displacement laws~\cite{Goldenberg02}.
\begin{figure*}
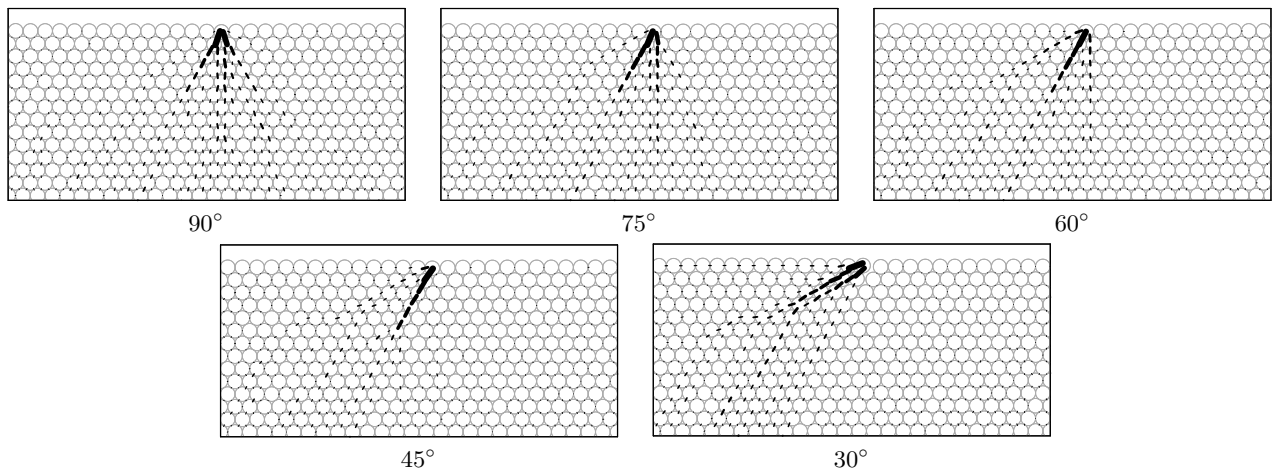

  \begin{center}
    \begin{tabular}{ccc}
      \includegraphics[width=2.1in]{chops04_gm070_v2_fig10.eps}&
      \includegraphics[width=2.1in]{chops04_gm070_v2_fig11.eps}&
      \includegraphics[width=2.1in]{chops04_gm070_v2_fig12.eps}\\
      $90^{\circ}$&$ 75^{\circ}$& $60^{\circ}$
    \end{tabular}
    \begin{tabular}{cc}
      \includegraphics[width=2.1in]{chops04_gm070_v2_fig13.eps}&
      \includegraphics[width=2.1in]{chops04_gm070_v2_fig14.eps}\\
      $45^{\circ}$&$30^{\circ}$
    \end{tabular}
    \vspace{-0.1in}
  \end{center}
  \caption{Force chains in 2D packings of slightly polydisperse frictional
    particles. A force $F$ with magnitude $150$ times the mean particle weight
    is applied to the particle at the center of the top layer. The angle of the
    force with respect to the horizontal is indicated below each picture. The
    same realization of the packing was used in all cases. The region shown is
    the central third of the upper half of the system.}
  \label{fig:forces_all_angles}
\end{figure*}

In order to elucidate the role of frictional forces and torques in the
quasi-static response of granular materials in general, and in particular in
order to gain an understanding of the experimental results mentioned
above~\cite{Geng03}, we performed discrete element simulations with normal and
tangential linear spring-dashpot forces among the particles (see
e.g.,~\cite{Cundall79,Herrmann98}), possibly the simplest model for frictional
disks. The simulation parameters were chosen to correspond to those of the
experimental system~\cite{GengPC}. Experimentally, the force-displacement law
for the photoelastic disks was found to be fit quite well by $f\propto
\xi^{3/2}$~\cite{GengPC}, as predicted by the standard Hertz theory for elastic
ellipsoids in contact (see e.g.~\cite{Landau86}), rather than the linear
relation (with logarithmic correction) expected for cylinders in
contact~\cite{Schwartz71}, which appears to imply that the contact region
between the ``disks'' is elliptic rather than rectangular. The simulation model
described above employs a linear force-displacement law, so that an effective
mean spring constant was estimated on the basis of the range of forces used in
the experiments.  The tangential spring constant was taken to be one-half the
normal spring constant (a rough estimate consistent with the Hertz-Mindlin
model~\cite{Mindlin53} for oblique contact forces). The normal and tangential
spring constants used are $k_n=3000\bar{m}g/\bar{R}$ and
$k_t=1500\bar{m}g/\bar{R}$, where $\bar{R}$ and $\bar{m}$ are the mean particle
radius and mass, respectively, and $g$ is the gravitational acceleration. The
friction coefficient used is $\mu=0.94$ for particle-particle contacts and
$\mu^{\rm wall}=0.35$ for particle-wall contacts. The systems studied here are
composed of polydisperse disks, with radii distributed uniformly in the
interval $[R-\delta R,R]$, where $\delta R/R=8\cdot 10^{-3}$ (i.e., a small
polydispersity).

The system is first relaxed to a static state under gravity (until the total
kinetic energy per particle is less than $10^{-9}\bar{m}g\bar{R}$), and then
relaxed again with an external force applied at the center of the top layer (in
some cases the force was increased linearly with time from zero to prevent the
``buckling'' of the top layer which leads to major rearrangements; these are
beyond the nearly elastic behavior considered here).

For comparison with the experiments presented in~\cite{Geng03}, an external
force of magnitude $F=150\bar{m}g$ was applied to the center top particle at
angles of $15^{\circ}, 30^{\circ}, 45^{\circ}, 60^{\circ}, 75^{\circ}$, and
$90^{\circ}$. The simulated systems consisted of $29$ rows of $80$ particles,
which is similar to the size of the systems used in the
experiments~\cite{Geng01,Geng03,GengPC}. Fig.~\ref{fig:forces_all_angles}
presents the forces obtained for different applied force angles.  The same
particle configuration was used in all cases.  No significant particle rotation
occurred except for the particles adjacent to the one on which the force is
applied. For a force at an angle of $15^{\circ}$, buckling occurred in the top
row, causing major rearrangements.  Such buckling was also observed in the
experiments, where it was apparently stabilized, limiting the rearrangements to
a small region near the point of application of the force, but we have not been
able to prevent major rearrangements in the simulation.  As mentioned,
tangential forces such as friction give rise to interparticle torques.
Simulations with an applied torque (in addition to the applied force) show that
this torque does influence the observed force chains~\cite{GoldenbergIP}.

The results are quite similar, qualitatively, to those observed in the
experiment~\cite{Geng03}. Note that the results shown in
Fig.~\ref{fig:forces_all_angles} are for a single configuration, while the
results presented in~\cite{Geng03} are for an average over configurations. The
results obtained in simulations for different realizations of the disorder are
qualitatively similar~\cite{GoldenbergIP}. The agreement of the results
obtained using a relatively simple force model with the experiments is
encouraging. A more detailed study of the effects of friction on the forces and
the stress field will be presented elsewhere~\cite{GoldenbergIP}.

\section{Concluding remarks}
\label{sec:conclusions}
We have shown that the seemingly inconsistent results of different kinds of
experiments studying the static response of granular packings to a
concentrated force can all be understood within the same framework of an
essentially elastic (elliptic) picture once the distinction between forces and
stress is made and the possible consequences of small system size, as well as
anisotropy, are taken into account. The effect of applied stresses on the
contact network may be modeled as a nonlinear, incrementally elastic model
(which may be further extended to describe yielding).

Somewhat surprisingly, many aspects of the response of such systems can be
understood using models of frictionless particles. However, some effects do
require the introduction of friction, as in the example of the force chains
obtained for oblique applied forces described in this paper. We note, however,
that the model for the friction used in the simulations described in
Sec.~\ref{sec:friction} consists of tangential springs (with the additional
Coulomb condition). This indicates that even for static frictional systems
(below yield) an elastic continuum model, which probably includes rotational
degrees of freedom (e.g., a Cosserat continuum model~\cite{Eringen68}), may be
appropriate.

Another important issue which requires further study is the effect of disorder,
and the relation of spatial averaging to averaging over the disorder.


\begin{thebibliography}{10}

\bibitem{Levy01}
A.~Levy and H.~Kalman, editors,
\newblock {\em Handbook of Conveying and Handling of Particulate Solids},
  Elsevier, 2001.

\bibitem{Jaeger96b}
H.~M. Jaeger, S.~R. Nagel, and R.~P. Behringer,
\newblock Physics Today {\bf 49}, 32 (1996).

\bibitem{Jaeger96a}
H.~M. Jaeger, S.~R. Nagel, and R.~P. Behringer,
\newblock Rev. Mod. Phys. {\bf 68}, 1259 (1996).

\bibitem{DeGennes99}
P.~G. de~Gennes,
\newblock Rev. Mod. Phys. {\bf 71}, S374 (1999).

\bibitem{Kadanoff99}
L.~P. Kadanoff,
\newblock Rev. Mod. Phys. {\bf 71}, 435 (1999).

\bibitem{Goldhirsch03}
I.~Goldhirsch,
\newblock Ann. Rev. Fluid Mech. {\bf 35}, 267 (2003).

\bibitem{Nedderman92}
R.~M. Nedderman,
\newblock {\em Statics and Kinematics of Granular Materials} (Cambridge
  University Press, 1992).

\bibitem{Savage98b}
S.~B. Savage,
\newblock Modeling and granular materials boundary value problems,
\newblock in {\em Proceedings of the NATO Advanced Study Institute on Physics
  of Dry Granular Media, Carg{\`e}se, France, September 15-26, 1997}, edited by
  H.~J. Herrmann, J.~P. Hovi, and S.~Luding, pp. 25--95, Kluwer, 1998.

\bibitem{Drescher72}
A.~Drescher and G.~de~Josselin~de Jong,
\newblock J. Mech. Phys. Solids {\bf 20}, 337 (1972).

\bibitem{Cundall79}
P.~A. Cundall and O.~D.~L. Strack,
\newblock Geotechnique {\bf 29}, 47 (1979).

\bibitem{Radjai99}
F.~Radjai, S.~Roux, and J.~J. Moreau,
\newblock Chaos {\bf 9}, 544 (1999).

\bibitem{Wittmer96}
J.~P. Wittmer, P.~Claudin, M.~E. Cates, and J.-P. Bouchaud,
\newblock Nature {\bf 382}, 336 (1996).

\bibitem{Bouchaud95}
J.-P. Bouchaud, M.~E. Cates, and P.~Claudin,
\newblock J. de Physique I {\bf 5}, 639 (1995).

\bibitem{Wittmer97}
J.~P. Wittmer, M.~E. Cates, and P.~Claudin,
\newblock J. de Physique I {\bf 7}, 39 (1997).

\bibitem{Tkachenko99}
A.~V. Tkachenko and T.~A. Witten,
\newblock Phys. Rev.~E {\bf 60}, 687 (1999).

\bibitem{Bouchaud01}
J.-P. Bouchaud, P.~Claudin, D.~Levine, and M.~Otto,
\newblock Eur. Phys. J.~E {\bf 4}, 451 (2001).

\bibitem{Socolar02}
J.~E.~S. Socolar, D.~G. Schaeffer, and P.~Claudin,
\newblock Eur. Phys. J.~E {\bf 7}, 353 (2002).

\bibitem{Otto03}
M.~Otto, J.-P. Bouchaud, P.~Claudin, and J.~E.~S. Socolar,
\newblock Phys. Rev.~E {\bf 67}, 031302 (2003).

\bibitem{Silva00}
M.~D. Silva and J.~Rajchenbach,
\newblock Nature {\bf 406}, 708 (2000).

\bibitem{Reydellet01}
G.~Reydellet and E.~Cl\'{e}ment,
\newblock Phys. Rev. Lett. {\bf 86}, 3308 (2001).

\bibitem{Serero01}
D.~Serero, G.~Reydellet, P.~Claudin, E.~Cl\'{e}ment, and D.~Levine,
\newblock Eur. Phys. J.~E {\bf 6}, 169 (2001).

\bibitem{Geng01}
J.~Geng {\em et~al.},
\newblock Phys. Rev. Lett. {\bf 87}, 035506 (2001).

\bibitem{Geng03}
J.~Geng, G.~Reydellet, E.~Cl\'{e}ment, and R.~P. Behringer,
\newblock Physica D {\bf 182}, 274 (2003).

\bibitem{Mueggenburg02}
N.~W. Mueggenburg, H.~M. Jaeger, and S.~R. Nagel,
\newblock Phys. Rev.~E {\bf 66}, 031304 (2002).

\bibitem{Moukarzel03}
C.~F. Moukarzel, H.~Pancheco-Mart\'{i}nez, J.~C. Ruiz-Suarez, and A.~M.
  Vidales,
\newblock cond-mat/0308240.

\bibitem{Goldenberg02}
C.~Goldenberg and I.~Goldhirsch,
\newblock Phys. Rev. Lett. {\bf 89}, 084302 (2002).

\bibitem{Goldhirsch02}
I.~Goldhirsch and C.~Goldenberg,
\newblock Eur. Phys. J.~E {\bf 9}, 245 (2002).

\bibitem{Tanguy02}
A.~Tanguy, J.~P. Wittmer, F.~Leonforte, and J.-L. Barrat,
\newblock Phys. Rev.~B {\bf 66}, 174205 (2002).

\bibitem{Moukarzel01}
C.~F. Moukarzel,
\newblock Granular Matter {\bf 3}, 41 (2001).

\bibitem{Head01}
D.~A. Head, A.~V. Tkachenko, and T.~A. Witten,
\newblock Eur. Phys. J.~E {\bf 6}, 99 (2001).

\bibitem{Silbert01}
L.~E. Silbert, D.~Erta\c{s}, G.~S. Grest, T.~C. Halsey, and D.~Levine,
\newblock Phys. Rev.~E {\bf 65}, 031304 (2001).

\bibitem{Herrmann98}
H.~J. Herrmann and S.~Luding,
\newblock Cont. Mech. and Thermodynamics {\bf 10}, 189 (1998).

\bibitem{Wolf96}
D.~E. Wolf,
\newblock Modeling and computer simulation of granular media,
\newblock in {\em Computational Physics}, edited by K.~H. Hoffmann and
  M.~Schreiber, pp. 64--94, Springer, Heidelberg, 1996.

\bibitem{Gladwell80}
G.~M.~L. Gladwell,
\newblock {\em Contact Problems in the Classical Theory of Elasticity}
  (Sijthoff \& Noordhoff, The Netherlands, 1980).

\bibitem{Johnson85}
K.~L. Johnson,
\newblock {\em Contact Mechanics} (Cambridge University Press, Cambridge,
  1985).

\bibitem{Landau86}
L.~Landau and E.~Lifshitz,
\newblock {\em Theory of Elasticity, 3rd Edition} (Pergamon, 1986).

\bibitem{Schafer96}
J.~Sch{\"a}fer, S.~Dippel, and D.~E. Wolf,
\newblock J. de Physique I {\bf 6}, 5 (1996).

\bibitem{Sadd93}
M.~H. Sadd, Q.~Tai, and A.~Shukla,
\newblock Int. J. of Non-Linear Mech. {\bf 28}, 251 (1993).

\bibitem{Walton95}
O.~R. Walton,
\newblock Force models for particle-dynamics simulations of granular materials,
\newblock in {\em Mobile Particulate Systems}, edited by E.~Guazzelli and
  L.~Oger, pp. 367--380, Kluwer, 1995.

\bibitem{Moukarzel98c}
C.~F. Moukarzel,
\newblock Phys. Rev. Lett. {\bf 81}, 1634 (1998).

\bibitem{Roux00}
J.-N. Roux,
\newblock Physical Review E {\bf 61}, 6802 (2000).

\bibitem{Makse00}
H.~Makse, D.~L. Johnson, and L.~M. Schwartz,
\newblock Phys. Rev. Lett. {\bf 84}, 4160 (2000).

\bibitem{Howell99}
D.~W. Howell and R.~P. Behringer,
\newblock Chaos {\bf 9}, 559 (1999).

\bibitem{Mueth98}
D.~M. Mueth, H.~M. Jaeger, and S.~R. Nagel,
\newblock Phys. Rev.~E {\bf 57}, 3164 (1998).

\bibitem{Blair01}
D.~L. Blair, N.~W. Mueggenburg, A.~H. Marshall, H.~M. Jaeger, and S.~R. Nagel,
\newblock Phys. Rev.~E {\bf 63}, 041304 (2001).

\bibitem{Radjai96}
F.~Radjai, M.~Jean, J.-J. Moreau, and S.~Roux,
\newblock Phys. Rev. Lett. {\bf 77}, 274 (1996).

\bibitem{Erikson02}
J.~M. Erikson, N.~W. Mueggenburg, H.~M. Jaeger, and S.~R. Nagel,
\newblock Phys. Rev.~E {\bf 66}, 040301 (2002).

\bibitem{Nguyen00}
M.~L. Nguyen and S.~N. Coppersmith,
\newblock Phys. Rev.~E {\bf 62}, 5248 (2000).

\bibitem{Silbert02a}
L.~E. Silbert, G.~S. Grest, and J.~W. Landry,
\newblock Phys. Rev.~E {\bf 66}, 061303 (2002).

\bibitem{O'hern00}
C.~S. O'Hern, S.~A. Langer, A.~J. Liu, and S.~R. Nagel,
\newblock Phys. Rev. Lett. {\bf 86}, 111 (2000).

\bibitem{Liu95}
C.-H. Liu {\em et~al.},
\newblock Science {\bf 269}, 513 (1995).

\bibitem{Coppersmith96}
S.~N. Coppersmith, C.-H. Liu, S.~Majumdar, O.~Narayan, and T.~A. Witten,
\newblock Phys. Rev.~E {\bf 53}, 4673 (1996).

\bibitem{O'hern02}
C.~S. O'Hern, S.~A. Langer, A.~J. Liu, and S.~R. Nagel,
\newblock Phys. Rev. Lett. {\bf 88}, 075507 (2002).

\bibitem{Glasser01}
B.~J. Glasser and I.~Goldhirsch,
\newblock Phys. Fluids {\bf 13}, 407 (2001).

\bibitem{Born88}
M.~Born and K.~Huang,
\newblock {\em Dynamical theory of crystal lattices} (Clarendon Press, Oxford,
  1988).

\bibitem{GoldenbergUP}
C.~Goldenberg and I.~Goldhirsch,
\newblock Unpublished.

\bibitem{Luding97}
S.~Luding,
\newblock Phys. Rev.~E {\bf 55}, 4720 (1997).

\bibitem{Cates99}
M.~E. Cates, J.~P. Wittmer, J.-P. Bouchaud, and P.~Claudin,
\newblock Chaos {\bf 9}, 511 (1999).

\bibitem{Oda93}
M.~Oda,
\newblock Mech. Mat. {\bf 16}, 35 (1993).

\bibitem{GengPC}
J.~Geng and R.~P. Behringer,
\newblock private communication  (2002).

\bibitem{Vu-Quoc99}
L.~Vu-Quoc and X.~Zhang,
\newblock Mech. Mat. {\bf 31}, 235 (1999).

\bibitem{Schwartz71}
J.~Schwartz and E.~Y. Harper,
\newblock Int. J. Solids and Structures {\bf 7}, 1613 (1971).

\bibitem{Mindlin53}
R.~D. Mindlin and H.~Deresiewicz,
\newblock J. Appl. Mech. {\bf 20}, 327 (1953).

\bibitem{GoldenbergIP}
C.~Goldenberg and I.~Goldhirsch,
\newblock In preparation.

\bibitem{Eringen68}
A.~C. Eringen,
\newblock Theory of micropolar elasticity,
\newblock in {\em Fracture: An Advanced Treatise, Volume {II}: Mathematical
  Fundamentals}, edited by H.~Liebowitz, pp. 621--729, Academic Press, New York
  and London, 1968.

\end{thebibliography}
\end{document}